\documentclass[12pt]{iopart}

\usepackage[version=4]{mhchem} 
\usepackage{xcolor} 
\usepackage{graphicx}
\usepackage{dcolumn}
\usepackage{bm}
\usepackage{tabularx}
\usepackage{makecell}
\usepackage{float}
\usepackage{multirow}
\usepackage{siunitx}
\usepackage{iopams}  

\begin{document}

\title{Investigations of the size distribution and magnetic properties of nanoparticles of Cu$_2$OSeO$_3$}

\author{S. J. R. Holt}
\ead{S.J.R.Holt@warwick.ac.uk}
\address{University of Warwick, Department of Physics, Coventry, CV4 7AL, United Kingdom}
\author{A. \v{S}tefan\v{c}i\v{c}}
\address{University of Warwick, Department of Physics, Coventry, CV4 7AL, United Kingdom}
\author{J. C. Loudon}
\address{University of Cambridge, Department of Materials Science and Metallurgy, Cambridge, CB3 0FS, United Kingdom}
\author{M. R. Lees}
\address{University of Warwick, Department of Physics, Coventry, CV4 7AL, United Kingdom}
\author{G. Balakrishnan}
\ead{G.Balakrishnan@warwick.ac.uk}
\address{University of Warwick, Department of Physics, Coventry, CV4 7AL, United Kingdom}

\begin{abstract}
Skyrmions in confined geometries have been a subject of increasing interest due to the different properties that they exhibit compared to their bulk counterparts.
In this study, nanoparticles of skyrmion-hosting $\text{Cu}_{2}\text{OSeO}_{3}$ have been synthesised using a precipitation method followed by thermal treatment.
This enables us to produce nanoparticles whose mean size varies from tens of nanometers to a few micrometers by varying the temperature and duration of the thermal decomposition of the precursor. 
These sizes span the $\sim 63$~nm diameter of skyrmions in $\text{Cu}_{2}\text{OSeO}_{3}$, allowing investigations into how the magnetic state changes when the size of the geometrical confinement is similar to and smaller than the size of an isolated magnetic skyrmion.
AC susceptibility measurements performed on nanoparticles with a size distribution from 15 to 250 nm show a change in the magnetic phase diagram compared to bulk $\text{Cu}_{2}\text{OSeO}_{3}$.
\end{abstract}


\maketitle

\section{Introduction}
Magnetic skyrmions are topologically stable magnetic spin textures which have properties of quasi-particles \cite{bogdanov1989thermodynamically}.
These magnetic vortex-like structures are of great interest to the research community, offering many potential uses in spintronic devices and for applications such as data storage \cite{nagaosa2013topological, fert2013skyrmions}.
Magnetic skyrmions have been observed in a variety of forms such as Bloch and N\'{e}el type skyrmions, both with a topological charge of $+1$ and a vorticity of $\pi/2$ and $0$ respectively \cite{muhlbauer2009skyrmion, kezsmarki2015neel}.
The counterpart to these are antiskyrmions with a topological charge of $-1$ which have been observed in magnets with D2d symmetry \cite{nayak2017magnetic, bogdanov1989thermodynamically}.
Alongside these skyrmions, there are a plethora of other predicted topologically non-trivial magnetic quasi-particles \cite{gobel2021beyond} such as higher-order skyrmions \cite{ozawa2017zero}, skyrmionium \cite{zhang2016control, zhang2018real}, bimerons \cite{kharkov2017bound, gobel2019magnetic, gao2019creation}, chiral bobbers \cite{rybakov2015new, zheng2018experimental}, and hopfions \cite{liu2018binding, sutcliffe2018hopfions}.
Some of this rich selection of magnetic phases have the potential to be stabilised in different magnetic materials.

Bloch-skyrmions in B20 chiral magnets manifest as a two-dimensional skyrmions that extend in a third dimension creating a tube like objects. 
Rather than being completely transitionally invariant along this third dimension, these skyrmion tubes are modulated near the surface of the sample in a way known as a chiral twist \cite{leonov2016chiral, wilson2014chiral, zhang2018reciprocal}.
This chiral twist occurs in the vicinity of the surface of the sample on a length scale comparable to that of the magnetic modulation length of the skyrmions, reducing the total energy of the system \cite{zhang2018reciprocal, rybakov2013three, wolf2021unveiling}.
At the surface of the skyrmion tube there is also a change in the angle of magnetisation known as surface twists which have an additional N\'eel contribution \cite{leonov2016chiral, zhang2018reciprocal}. 
These edge effects allow for the stabilisation of particle-like states such as chiral bobbers which manifest in a similar way to a skyrmion tube, but rather than the tube propagating all the way through the material, they end in a Bloch point \cite{rybakov2015new, zheng2018experimental}.
These chiral bobbers have a characteristic depth which is less than half of the magnetic modulation length and can be stabilised in a wide range of magnetic phase diagrams, highlighting the importance of surface effects along the axis of skyrmion tubes.
There are also additional confinement effects originating from the interaction of the samples edge with the skyrmion tube.
Often, skyrmions form a hexagonal lattice, in a region just below the Curie temperature $T_\textrm{C}$, but due to their topological nature, skyrmions can also be stabilised as isolated quasi-particles \cite{muhlbauer2009skyrmion}.
Skyrmions have a repulsive interaction with the edge of the sample \cite{iwasaki2013current} leading to the confinement of skyrmions within a sample.
In order to study these confinement effects the size of the particles in which the skyrmions are stabilised have to be comparable to the average size or modulation length of the skyrmions in these materials, which has been well established in most bulk materials.
One such route to study the confinement effects is in the investigation of the behaviour of nanoparticles of skyrmion hosts, making nanoparticles a prime contender for the confinement of magnetic skyrmions.

Skyrmions in B20 materials have a diameter determined by the material parameters of the Heisenberg exchange and Dzyaloshinskii–Moriya interactions, these sizes are typically on the scale of tens of nanometers.
For example, Bloch-type skyrmions have been experimentally identified in B20 materials such as MnSi with a size of 17~nm, FeGe with a size of 70~nm, and Cu$_2$OSeO$_3$ with a size of 63~nm ~\cite{muhlbauer2009skyrmion, yu2011near, seki2012observation}.
It has been observed that physically changing the size and shape of a material can dramatically change the region of the phase diagram over which of skyrmions are stabilised.
For example, in Lorentz transmission electron microscopy (LTEM) studies, thinning the sample dramatically increases the range of temperatures and magnetic fields over which skyrmions are stable~\cite{seki2012observation}.
This highlights the considerable effects that confinement along the axis of a skyrmion tube have on the stabilisation of skyrmions.
In MnSi, investigations have demonstrated the effect of confinement on skyrmions by comparing thin films~\cite{tonomura2012real}, nanowires~\cite{yu2013observation}, nanoparticles~\cite{das2018effect}, and bulk~\cite{muhlbauer2009skyrmion}.
Nanowires exhibit the largest enhancement to the extent of the skyrmion phase in the magnetic phase diagram.
The skyrmion phase in thin films is found at a reduced temperature and higher magnetic fields compared to the bulk.
In nanoparticles, previous observations indicate an increase in the temperature at which skyrmions are observed, making them stable only under very small fields~\cite{das2018effect}.

In this study, we focus on the skyrmion-hosting material Cu$_2$OSeO$_3$, and the effects of three-dimensional geometric confinement of the skyrmions in nanoparticles of this material.
Cu$_2$OSeO$_3$ is a multiferroic insulator that has been investigated extensively for multiple reasons including the use of an external electric field to manipulate skyrmions~\cite{white2012electric, seki2012observation}, exhibiting two distinct skyrmion phases~\cite{chacon2018observation}, and the ability to study the effects of chemically substitution of Cu atoms with Zn and Ag~\cite{stefancic2018origin, neves2020effect}.
Cu$_2$OSeO$_3$ crystallises in the chiral $P2_1 3$ space group ($T$ point group), a symmetry group in which the majority of known skyrmion materials have been found~\cite{tokura2020magnetic}.
There are two crystallographically inequivalent Cu sites present in a $1:3$ ratio with a spin $1/2$ pointing in opposite directions on each site, creating a ferrimagnetic lattice~\cite{bos2008magnetoelectric, belesi2010ferrimagnetism, maisuradze2011mu} where superexchange via the O atoms mediates the magnetic interactions~\cite{yang2012strong, vzivkovic2012two}.

Recently, there have been micromagnetic investigations into the confinement of skyrmions in nanoparticles of FeGe~\cite{pathak2021three}.
Due to the similar magnetic modulation period of both FeGe and Cu$_2$OSeO$_3$, micromagnetic simulations of FeGe also describe the ferrimagnetic Cu$_2$OSeO$_3$ system well~\cite{takagi2020particle}.
This means that results from micromagnetic studies on nanoparticles in FeGe systems are relevant to \ce{Cu2OSeO3} nanoparticles.
Pathak and Hertel~\cite{pathak2021three} found that the magnetic phases such as skyrmions, helices, merons, chiral bobbers, and saturated states can be stabilised across a variety of nanoparticle sizes under different applied magnetic fields compared to the bulk system.
For example, the skyrmion phase in a small applied magnetic field has been shown to give way to a meron phase and finally a saturated magnetic state upon reduction of the nanoparticle size.
Our study sets out to investigate if any changes in the magnetic state can be experimentally observed in nanoparticles of Cu$_2$OSeO$_3$ as the size of the nanoparticle approaches, and is reduced smaller than the size of an isolated magnetic skyrmion.

\section{Experimental Details}
Powder X-ray diffraction (PXRD) was performed on a Panalytical X-Pert Pro diffractometer operating in Bragg-Brentano geometry equipped with a monochromatic Cu K$_{\alpha 1}$ source and a solid-state PIXcel one-dimensional detector.
A Zeiss SUPRA 55-VP scanning electron microscope with an accelerating voltage of 10~kV was used for scanning electron microscopy (SEM) experiments.
Transmission electron microscopy (TEM) was performed using a FEI Titan electron microscope. 
A Mastersizer 2000 particle size analyser was used to measure the size of the \ce{Cu2OSeO3} nanoparticles by laser diffraction.
AC susceptibility measurements were performed using a Quantum Design Magnetic Property Measurement System (MPMS5) superconducting quantum interference device (SQuID) magnetometer as a function of DC applied field.
These AC susceptibility measurements were performed following a zero-field-cooling procedure with an AC field of 10~Hz and 0.3~mT.

\section{Sample Preparation}
\ce{Cu2OSeO3} nanoparticles were prepared using a precipitation method followed by thermal treatment~\cite{fokina2014thermodynamics}.
Copper(II) chloride dihydrate \ce{(CuCl2.2H2O)} and sodium selenite decahydrate \ce{(Na2SeO3.10 H2O)} were dissolved separately in distilled water with a small amount of \ce{HCl} added to the \ce{CuCl2} solution to reduce the pH.
The \ce{CuCl2} solution was slowly added to the \ce{Na2SeO3} solution producing a precipitate which gradually changed from milky green to blue upon adding additional \ce{CuCl2}.
This precipitate, which forms via the reaction
\begin{align}
\begin{split}
  \ce{ Na2SeO3_{(aq)} + & CuCl2_{(aq)} -> } \\
  \ce{& 2NaCl_{(aq)} + CuSeO3.2H2O_{(s)}}
\end{split}
\end{align}
was confirmed to be \ce{CuSeO3.2H2O} using powder X-ray diffraction.
The solid \ce{CuSeO3.2H2O} precipitate was filtered from the solution and washed with distilled water, isopropanol, and acetone before drying at 50~$^\circ$C on air for 18 hours.

\ce{Cu2OSeO3} nanoparticles were synthesised by heating \ce{CuSeO3.2H2O} in an alumina crucible under a flow of \ce{O2} in a tube furnace.
At the initial heating stage, a dehydration process takes place in multiple steps leading to the overall reaction~\cite{malta2019synthesis}
\begin{equation} 
    \ce{ CuSeO3.2H2O ->[\Delta T] CuSeO3 + 2H2O ^}.
\end{equation}
A thermal decomposition then takes place to convert the \ce{CuSeO3} into \ce{Cu2OSeO3} nanoparticles via the following reactions:
\begin{align}
    \ce{ 4CuSeO3 ->[\Delta T] Cu4O(SeO3)3 + SeO2 ^}, \\
    \ce{ Cu4O(SeO3)3 ->[\Delta T] 2Cu2OSeO3 + SeO2 ^}.
\end{align}
One of the key stages in the process is the thermal decomposition of \ce{Cu4O(SeO3)3} to \ce{Cu2OSeO3} as during this process the material goes through an amorphous phase~\cite{fokina2014thermodynamics}. 
The temperatures used here for the decomposition varied between 400~$^\circ$C and 470~$^\circ$C depending on the particle size desired. Fours sets of nanoparticles, hereafter referred to as Samples A, B, C, and D, were produced. 
The procedures used and the size of the nanoparticles produced (determined in the following sections) are detailed in Table~\ref{tab:size}.
It is worth noting that at higher temperatures or long time treatments, special care has to be taken not to thermally decompose the \ce{Cu2OSeO3} further via~\cite{malta2019synthesis}
\begin{equation} 
    \ce{ 2Cu2OSeO3 ->[\Delta T] 4CuO + 2SeO2 ^}.
\end{equation}

\noindent For comparison, a polycrystalline sample of $\text{Cu}_{2}\text{OSeO}_{3}$ (Sample SS) was synthesised by thoroughly grinding together stoichiometric amounts of CuO ($99.99\%$, metals basis, Alfa Aesar), $\text{SeO}_2$ ($99.999\%$, trace metal basis, Acros Organics) inside an argon-filled glove box.
The mixtures of powders were transferred into silica tubes, evacuated and sealed. 
Samples were then heated at a rate of 3.5~$^{\circ}$C/h to 650~$^{\circ}\text{C}$, kept at this temperature for 96~h, followed by water quench cooling.

\begin{table*}
    \centering
    \caption{Synthesis conditions of samples and sizes of \ce{Cu2OSeO3} particles. Samples A - D were prepared by precipitation and thermal treatment. Sample SS was prepared via a solid state reaction.} 
    \label{tab:size}
    \begin{tabular}{ccccc}
    \br
   Sample  & Temperature & Dwell time &\multicolumn{2}{c}{Particle diameter} \\
      &  $(^{\circ}\text{C})$ &  (h) & PXRD & SEM/TEM\\
    \mr
    A  & 400 then 420 & 288 then 48 &  $\sim$ \SI{3.5}{\micro\metre} & 1-8~\SI{}{\micro\metre} \\
    B  & 420 & 72 &  $\sim$ \SI{3.2}{\micro\metre} & 0.8-2~\SI{}{\micro\metre} \\
    C  & 450 & 24 &  $\sim 112$~nm & 15-250~nm \\
    D  & 470 & 24 &  $\sim 126$~nm & - \\
    \\
    SS  & 650 & 96 & - & 10-100~\SI{}{\micro\metre}\\
    \br
    \end{tabular}
\end{table*}

\section{Characterisation}
\subsection{Powder X-ray Diffraction} \label{sec:C6:PXRD}
\begin{figure}
    \centering
    \includegraphics[width=0.9\linewidth]{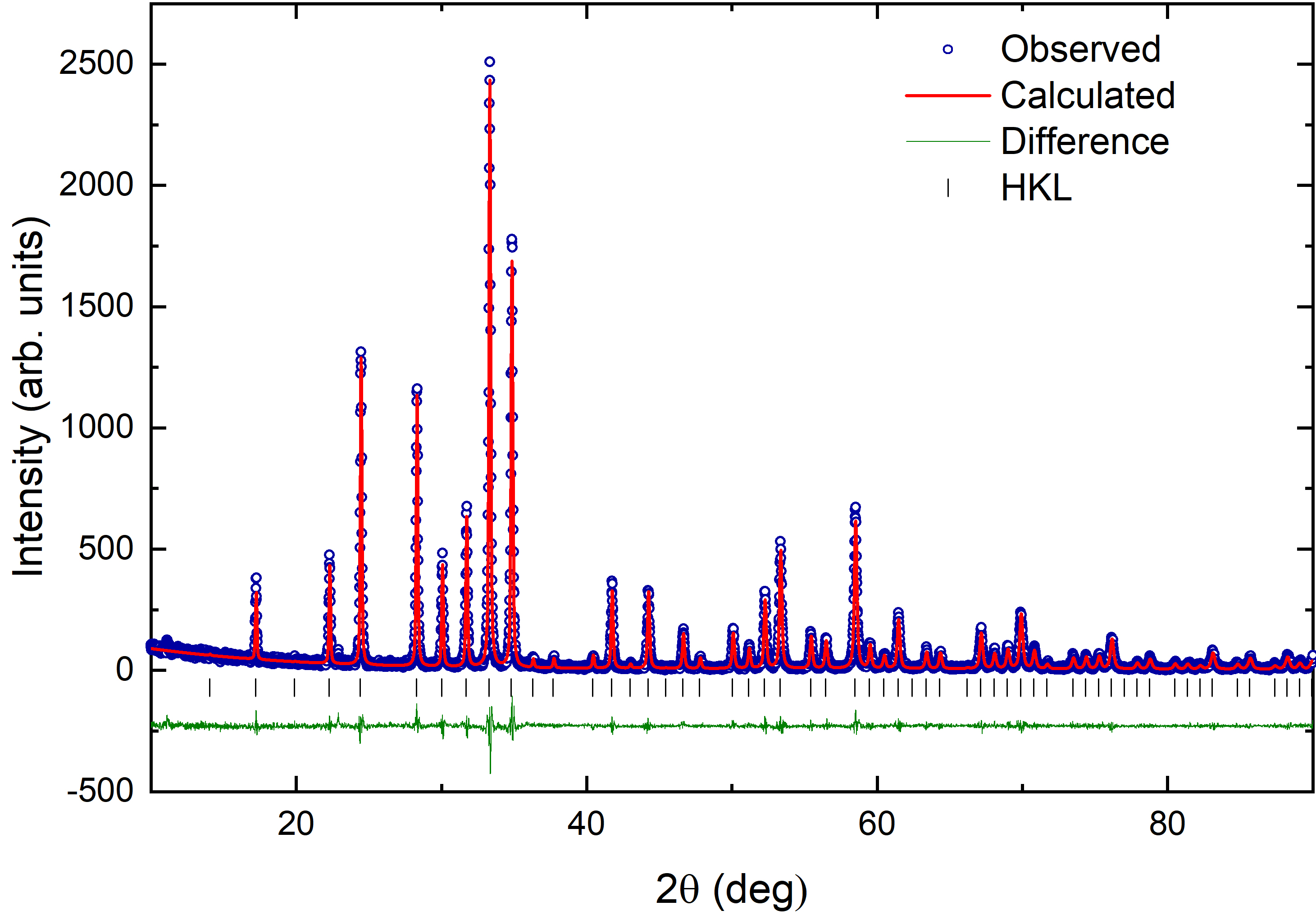}
    \caption{Powder X-ray diffraction profile of \ce{Cu2OSeO3} nanoparticles (Sample D).
    The experimentally-obtained diffraction profile at ambient temperature (blue open circles), refinement based on the model obtained from single crystal X-ray diffraction at room temperature (red solid line), difference (olive green solid line) and predicted peak positions (black tick marks).}
    \label{fig:C6:PXRD}
\end{figure}

The phase purity of the \ce{Cu2OSeO3} nanoparticles was examined by powder X-ray diffraction. All peaks could be indexed in the $P2_13$ space group, in excellent agreement with the literature~\cite{stefancic2018origin}.
Figure~\ref{fig:C6:PXRD} shows an X-ray diffraction pattern of \ce{Cu2OSeO3} nanoparticles from Sample D.
By measuring a Si standard, the instrumental broadening of the diffraction peaks can be measured, and hence the sample broadening can be isolated so that the crystallite domain size and the strain can be quantified using the Debye-Scherrer equation~\cite{scherrer1918nachr}.

\begin{figure}
    \centering
    \includegraphics[width=0.85\linewidth]{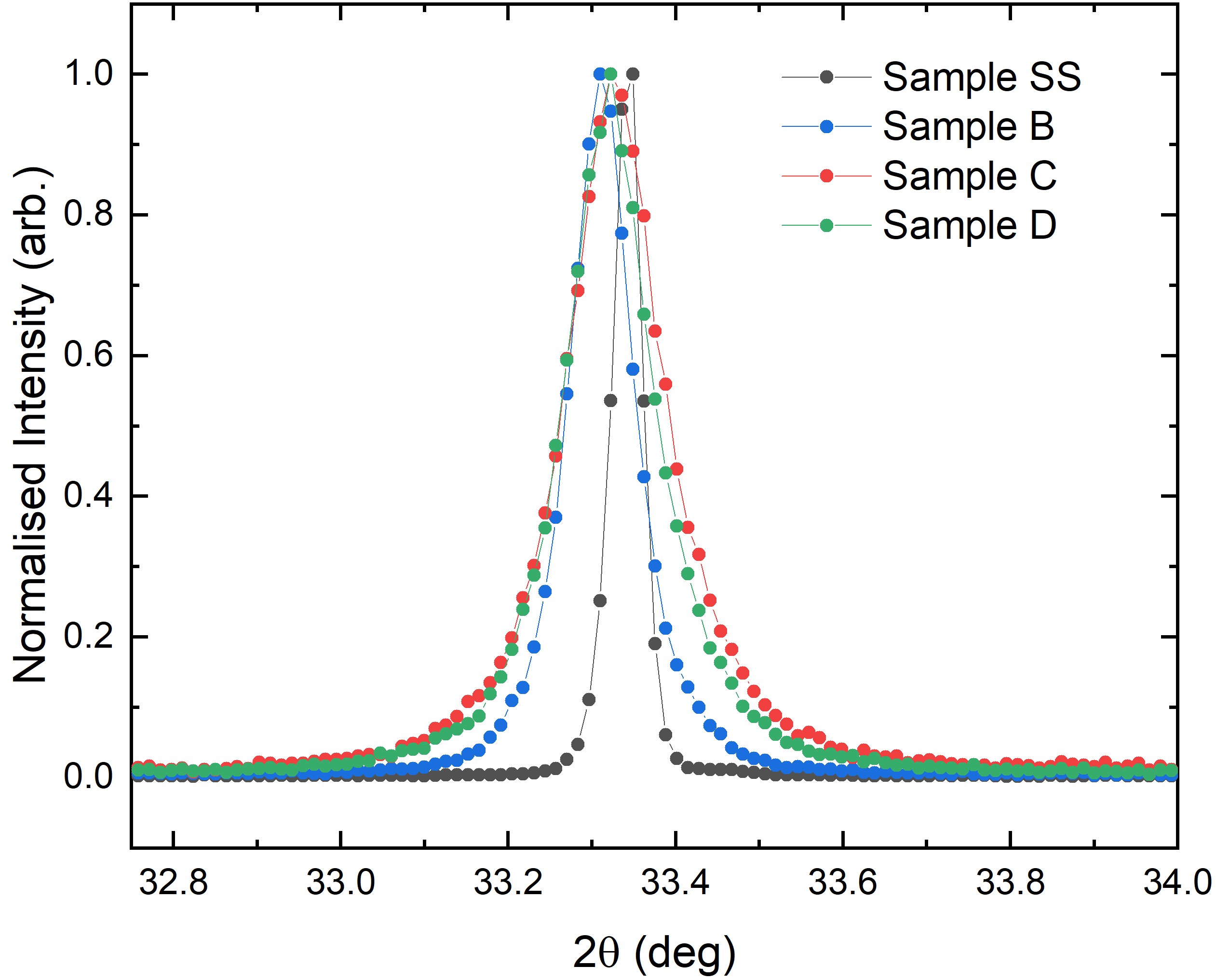}
    \caption{Powder X-ray diffraction of the (311) peak comparing \ce{Cu2OSeO3} produced by solid synthesis (Sample SS), and Samples B, C, and D obtained by the precipitation method followed by thermal treatment.}
    \label{fig:C6:PXRD_comp}
\end{figure}

Figure~\ref{fig:C6:PXRD_comp} shows a comparison of Sample SS, \ce{Cu2OSeO3} powder made using solid state synthesis, and nanoparticles from Samples B, C, and D synthesised via the precipitation method followed by thermal treatment.
A clear broadening can be seen in the Bragg peaks of the nanoparticles of \ce{Cu2OSeO3} compared to the powder prepared by solid state synthesis. This indicates that the nanoparticles have a smaller crystallite size.
If it is assumed that the crystallites are spherical, with a size distribution that follows a log-normal distribution~\cite{popa202analytical}, the volume-weighted domain size can be related to the radii of the crystallites~\cite{balzar2004size}.
Quantification of the crystallite sizes using Rietveld analysis can be seen in Table~\ref{tab:size}.

\subsection{Scanning Electron Microscopy}
\begin{figure}
\centering
\includegraphics[width=0.75\linewidth]{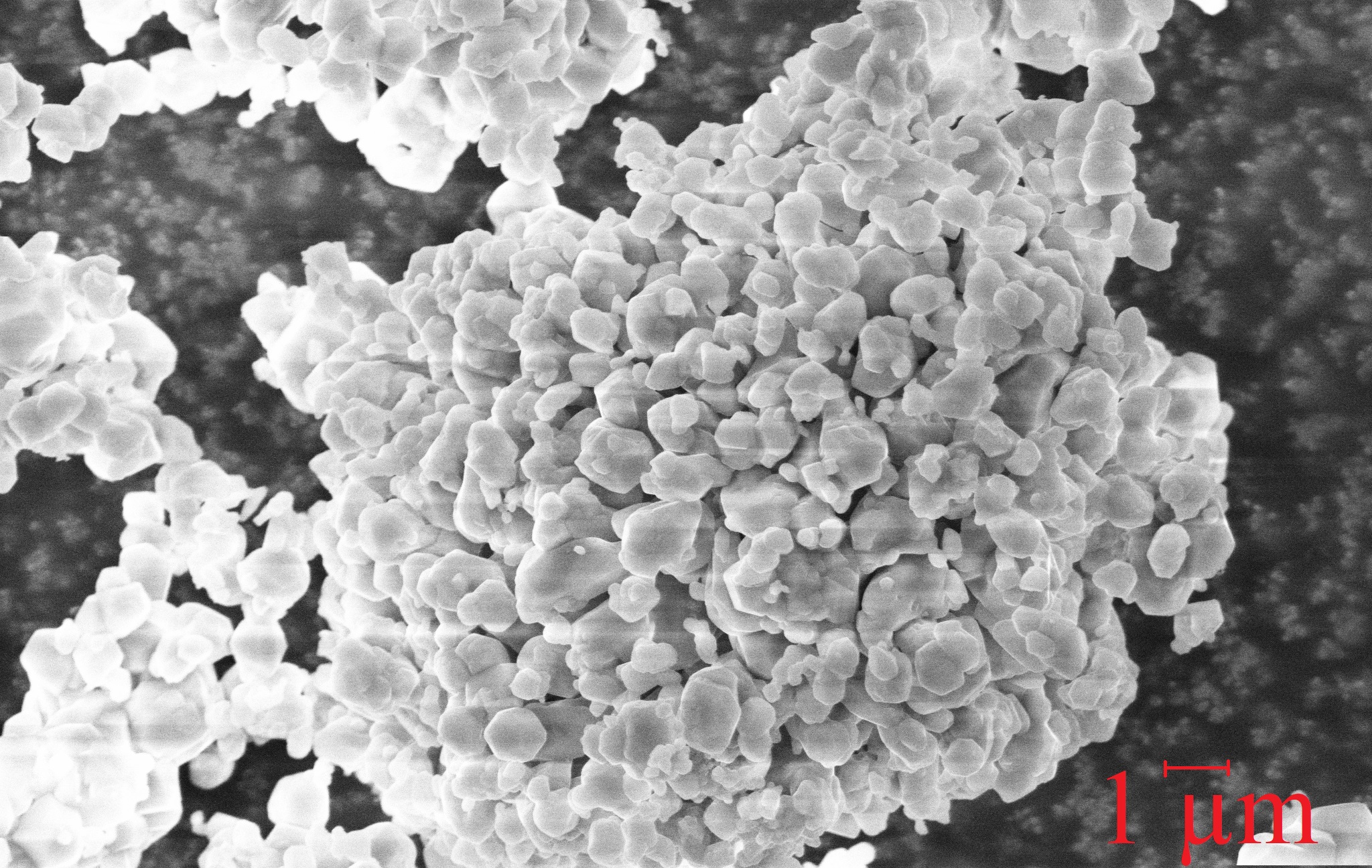}
\includegraphics[width=0.75\linewidth]{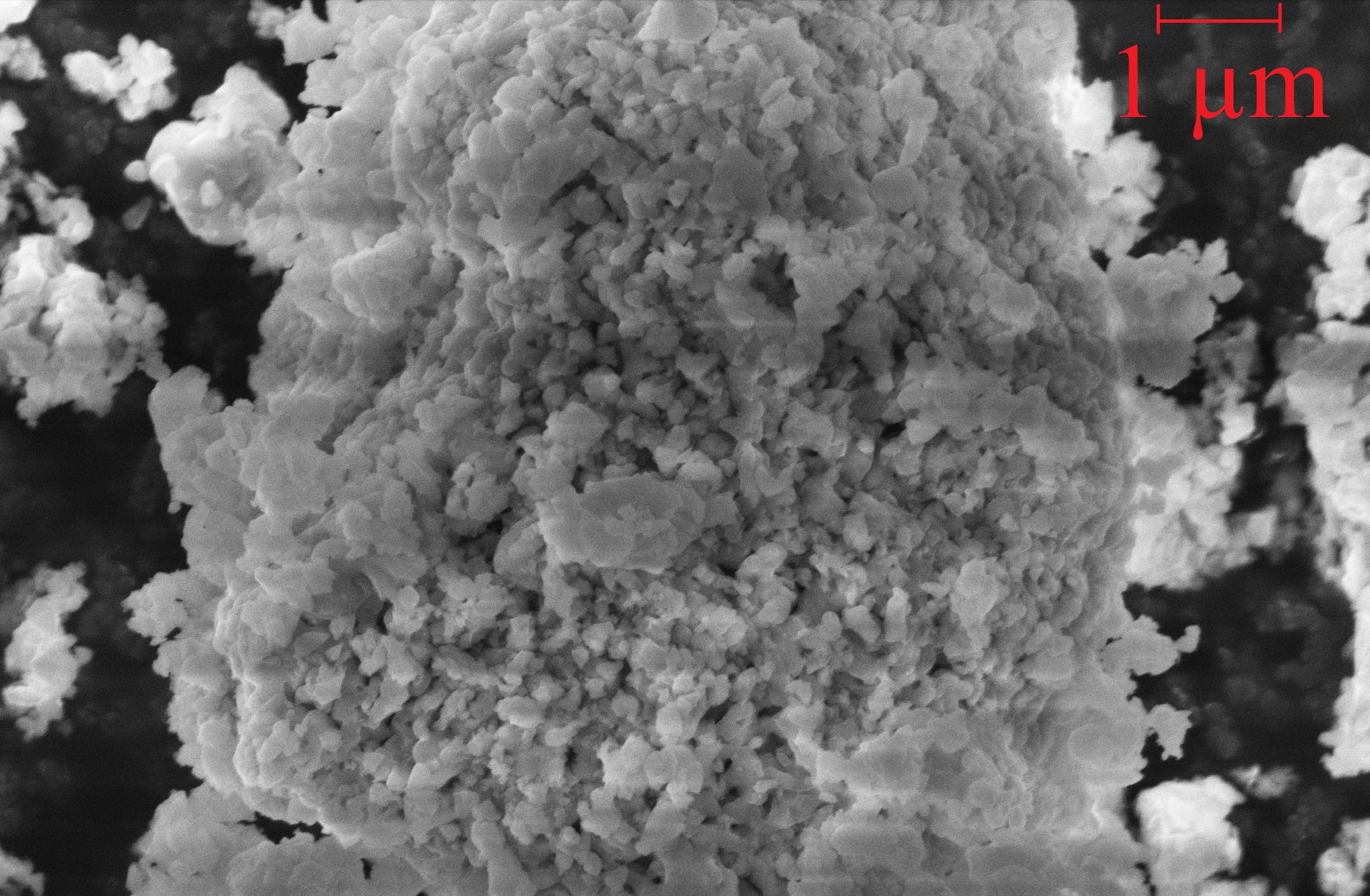}
\caption{SEM images of \ce{Cu2OSeO3} nanoparticles from (a) Sample B and (b) Sample C. The sizes of the individual particles and aggregates can be clearly seen.}
\label{Fig:C6:SEM}
\end{figure}

Scanning electron microscopy was used to image the particles in order to investigate their size and shape.
Table~\ref{tab:size} details the sizes of nanoparticles imaged by SEM.
Figure~\ref{Fig:C6:SEM} shows the SEM images of nanoparticles from Samples B and C.
It can be seen that there is a size distribution of particles which have an average diameter of $\sim 1$~\SI{}{\micro\metre} in Sample B, in agreement with the powder X-ray diffraction measurements.
The particles in Sample C can be seen to be significantly smaller with most particles ranging in diameter from 60 to 250~nm.
In both samples the particles are quite rough and tend to be seen in larger aggregates which can be up to $\sim 20$~\SI{}{\micro\metre} in diameter.

\subsection{Laser diffraction}
\begin{figure}
    \centering
    \includegraphics[width=0.95\linewidth]{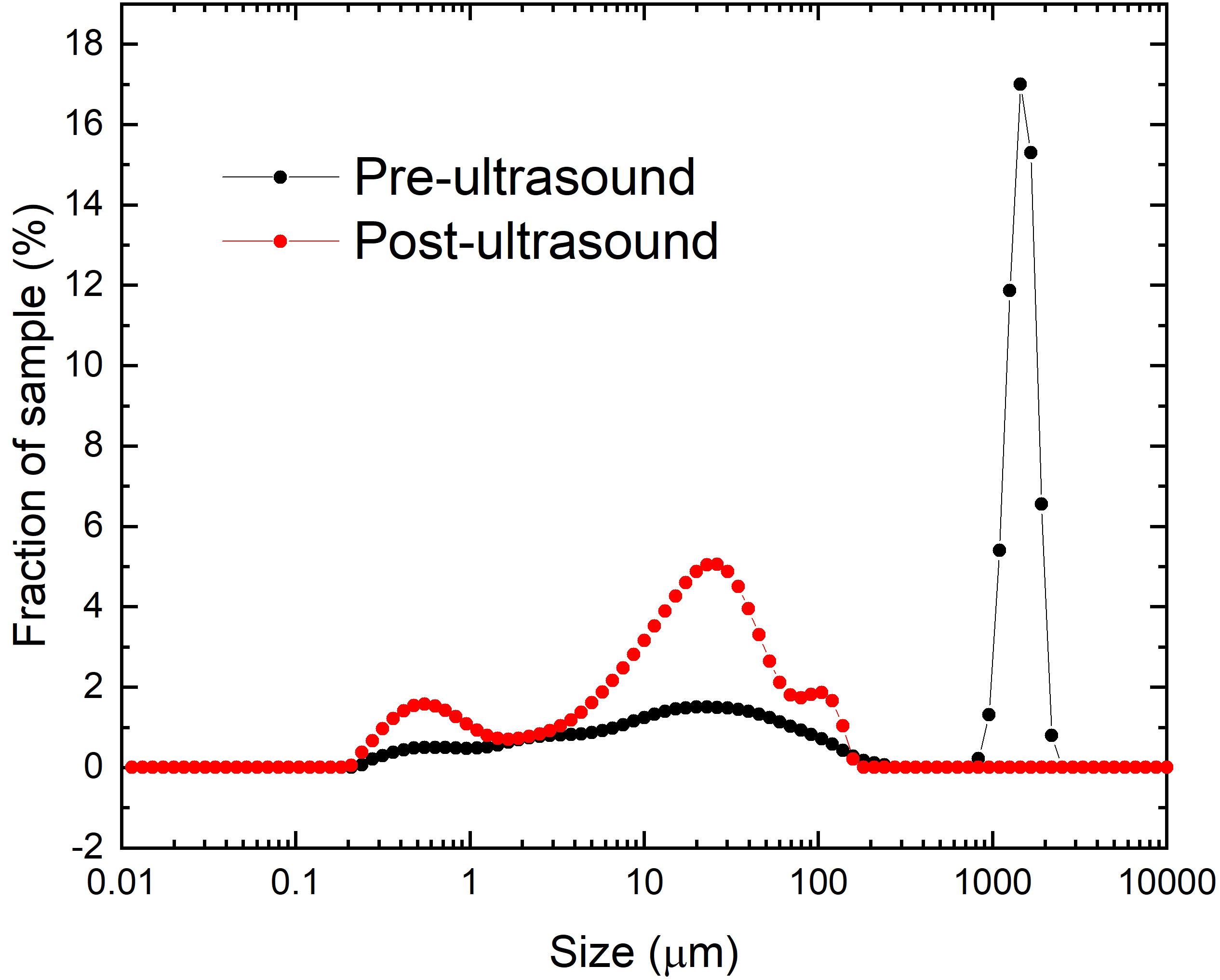}
    \caption{Particle size distribution of \ce{Cu2OSeO3} nanoparticles in Sample C determined from laser diffraction. The ultrasonication process has broken up the larger aggregates.}
    \label{Fig:C6:PS_1}
\end{figure}

Laser diffraction was used to measure the particle size in the powders.
This technique is based on the optical size of the object, hence aggregates appear as a larger particle rather than collections of smaller particles.
Figure~\ref{Fig:C6:PS_1} shows the measured size distributions for \ce{Cu2OSeO3} nanoparticles from Sample C suspended in distilled water.
With the application of ultrasound, the larger aggregates with a peak at $\sim 1500$~\SI{}{\micro\metre} break up into particles with sizes below $\sim 100$~\SI{}{\micro\metre}. 
These smaller particle sizes are similar to what is observed using TEM and SEM.
However, it can be seen that even with ultrasonication the measured size does not approach the crystallite size measured with X-ray diffraction or SEM, indicating that the majority of the nanoparticles remain in the form of aggregates.
Similar behaviour is seen for all of the \ce{Cu2OSeO3} nanoparticles synthesised.

\subsection{Transmission electron microscopy}
\begin{figure}
\centering
\includegraphics[width=0.75\linewidth]{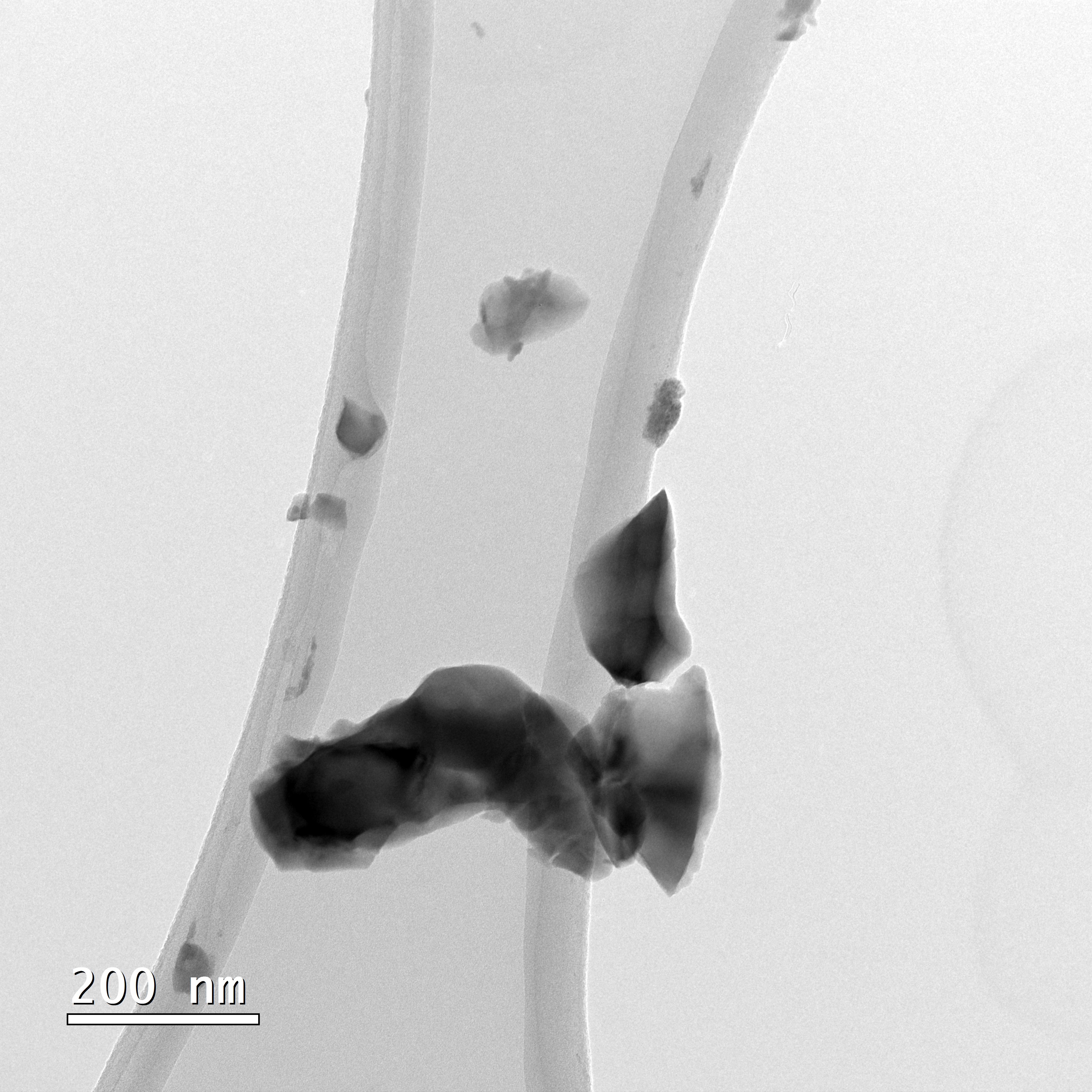}
\caption{TEM image of the \ce{Cu2OSeO3} nanoparticles in Sample C highlighting their irregular shapes and sizes.}
\label{Fig:C6:TEM}
\end{figure}

Transmission electron microscopy was used to image the nanoparticles to gain more information on their size and shape.
Nanoparticles were suspended in methanol and then subject to ultrasonication to disperse the nanoparticles and break up some of the larger aggregates.
A small part of the solution was then placed on a lacey carbon film.
Table~\ref{tab:size} details the sizes of nanoparticles imaged by TEM.
Figure~\ref{Fig:C6:TEM} shows a TEM image of the nanoparticles (Sample C).
There are a variety of different diameters of nanoparticles ranging from $\sim 15$ to $\sim 250$~nm.
Most of the particles are relatively rough as opposed to having well-defined crystallographic faces.
The sizes of these particles are in the range of interest, as they span the size of a skyrmion ($\sim 63$~nm) in \ce{Cu2OSeO3}. 

\section{Magnetism}
\subsection{AC susceptibility}
AC magnetometry was performed on nanoparticles of Samples C and D as these had the smallest average size, and the results were compared with measurements on Sample SS, in order to observe the differences in the magnetic phases exhibited by the bulk sample and the nanoparticles.
AC magnetometry is sensitive to the dynamics of the magnetism in the sample. If a different magnetic state is realised this may result in a change in the AC susceptibility response.

\begin{figure}
\centering
\includegraphics[width=1\linewidth]{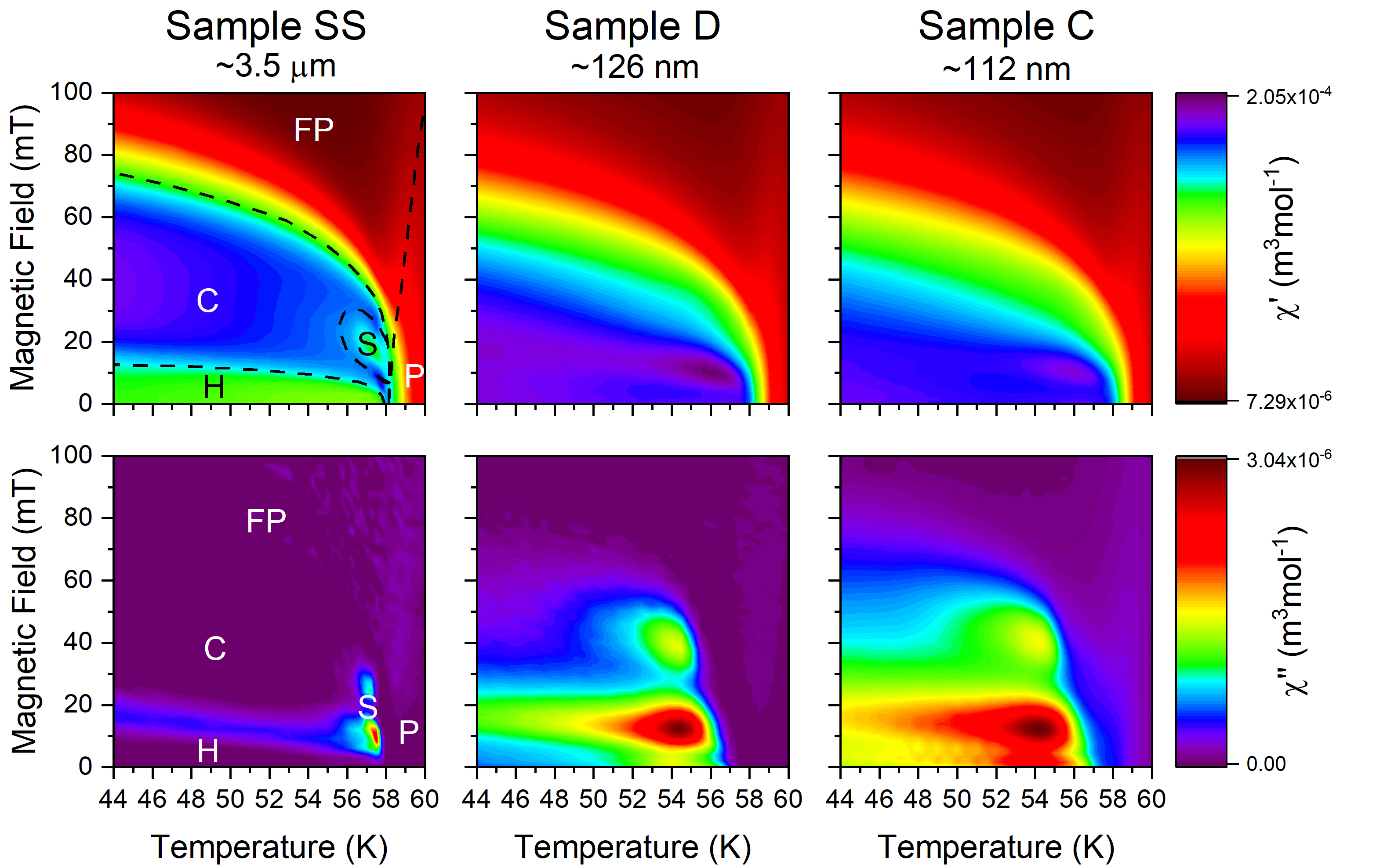}
\caption{(a-c) In-phase and (d-f) out-of-phase AC susceptibility as a function of temperature and applied DC field for polycrystalline (Sample SS) and nanopaticles (Samples C and D) of \ce{Cu2OSeO3}. The magnetic phases are given for polycrystalline \ce{Cu2OSeO3} where H is the helical phase, C is the conical phase, and S is the skyrmion lattice. FP and P are the field polarised and paramagnetic states, respectively.
The lines are given as a guide to the eye.
The crystallite sizes indicated are obtained from PXRD (see Table 1).
}
\label{Fig:C6:AC_maps}
\end{figure}

Figure~\ref{Fig:C6:AC_maps} shows the in-phase, $\chi'$, and out-of-phase component, $\chi''$, of the AC susceptibility as a function of temperature and magnetic feld for the nanoparticles (Samples C and D), compared to polycrystalline \ce{Cu2OSeO3} produced by solid state synthesis (Sample SS).
For $\chi'$, a clear difference can be seen in the phase diagrams, as the pocket associated with skyrmions in the polycrystalline \ce{Cu2OSeO3} is no longer visible for the nanoparticle samples.
The signals associated with the conical and helical phase are also dramatically changed as there are no longer clear plateaus that are indicative of both of these phases.

A comparison of the phase diagrams obtained from $\chi''$ for the different particles shows that there are some similar features present in all of them.
For the polycrystalline \ce{Cu2OSeO3} there is a peak just below 58~K associated with the transition to a skyrmion phase with a long ridge stretching from 58~K downwards linked to the helical to conical transition.
For the nanoparticles, the peak in $\chi''$ has shifted to $\sim 54$~K and a slightly higher DC magnetic field. The signal is also larger in magnitude, indicating higher dissipation. A ridge in $\chi''$ can also be seen in Samples C and D that extends from the peak at $\sim 54$~K downward in temperature.

\begin{figure}
\centering
\includegraphics[width=0.6\linewidth]{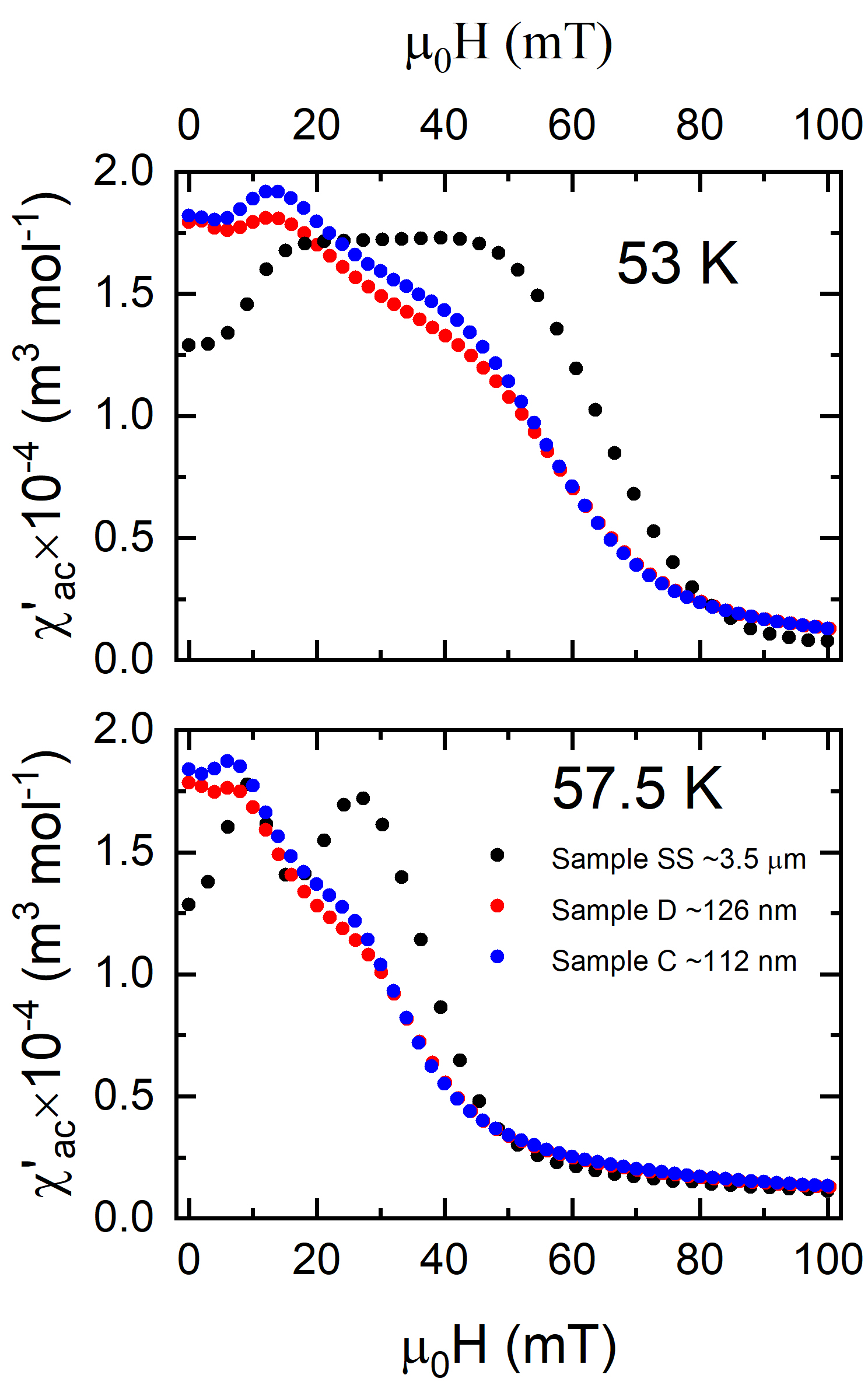}
\caption{In-phase AC susceptibility $\chi'$ versus applied DC magnetic field at 53~K and 57.5~K for polycrystalline \ce{Cu2OSeO3} and nanoparticles of \ce{Cu2OSeO3} taken from Sample C and D. The crystallite sizes indicated are obtained from PXRD.}
\label{Fig:C6:AC_53K}
\end{figure}

Figure~\ref{Fig:C6:AC_53K} shows an AC susceptibility field sweep that compares the nanoparticles with polycrystalline \ce{Cu2OSeO3}.
It is clear that the plateau of the helical phase in the polycrystalline material is no longer present in the nanoparticles and the magnitude of the susceptibility is higher compared to the magnitude of the conical phase in the polycrystalline \ce{Cu2OSeO3}.
However, there is a maximum at $\sim 12$~mT in $\chi'$ for the nanoparticles that occurs at the same point as the helical to conical transition.
There is also a shoulder at $\sim 45$~mT that is close to the conical to field polarised transition seen in polycrystalline \ce{Cu2OSeO3}.
At 57.5~K the features seen in the nanoparticles occur at the same fields as the peaks surrounding the skyrmion pocket at $\sim 8.5$~mT and $\sim 26$~mT.
These features suggest that these or similar transitions occur in at least some of the nanoparticles.
\section{Discussion and Summary}
Nanoparticles of \ce{Cu2OSeO3} have been synthesised via a precipitation method followed by thermal treatment with a degree of control over the size of nanoparticles produced.
X-ray diffraction was used to confirm the phase purity of the synthesised nanoparticles and the broadening of the diffraction peaks was used to calculate the crystallite sizes.
A clear correlation between particle size and synthesis conditions have been observed with thermal decomposition taking place at a higher temperature for a shorter amount of time producing smaller nanoparticles.
The ability to tune the nanoparticle crystallite size is likely due to the amortisation of the intermediate phase that occurs during the thermal decomposition process.

The particle shape and size distribution was also investigated using SEM and TEM, which corroborated the link between nanoparticle size and the synthesis conditions.
However, these techniques also revealed that the nanoparticles tend have rough surfaces and form in larger aggregates.
In addition, laser diffraction was used to determine the particle size distribution, but due to this technique being based on the optical size of particles rather than the crystallite size, it yielded results pointing to sizes that were larger than the actual nanoparticle size.
This means that this technique is limited to measuring the size of the aggregates rather than the size of the nanoparticles.
In addition, the size distribution produced with laser diffraction was approximately consistent for all of the samples produced by precipitation method followed by thermal treatment.

AC susceptibility was performed to examine the magnetic phases present. It is not possible to unambiguously identify a magnetic phase based on the response from AC susceptibility alone so either further measurements using different techniques such as electron holography are needed to identify magnetic phases or a comparison has to be made to.
Comparing the AC susceptibility phase diagrams of the nanoparticles with that of polycrystalline \ce{Cu2OSeO3}, where the magnetic phases are known it appears that there are significant changes in the magnetic states stabilised.
There are, howvever,  some important some similarities between in both the polycrystalline and nanoparticles AC susceptibility phase diagrams. 
For example, $\chi''$ for the nanoparticles has a set of peaks at 54~K.
This could be an indication that skyrmions still form in some of the nanoparticles, but are suppressed to a lower temperature.
A prominent ridge is also observed in $\chi''$ for the nanoparticles.
There are several possible causes of the ridge. It could be due to a helical to conical phase transition, as in the bulk material.
However, the ridge could also be due to a correlation between the temperature at which a peak in $\chi''$ appears and the crystallite size.
This would mean a distribution of particles sizes will cause a smearing of a peak into a ridge.

The investigations presented here clearly show the ability to tune the size of the nanoparticles formed depending on the thermal decomposition temperatures and dwell times.
It has also been suggested that a ferrimagnetic ground state is stabilised in nanoparticles of size smaller than the size of a skyrmion ~\cite{devi2019limit}.
While it is not possible from the investigation presented here to unequivocally confirm this, the AC susceptibility does suggest a distinct change in the magnetic phases present depending on the size of nanoparticles.
This is corroborated by micromagnetic simulations on a similar material, FeGe ~\cite{pathak2021three}.

While definite changes to the magnetism within nanoparticles have been observed, further experimental investigations of the effects on the magnetic structure of the nanoparticle size are required. 
If \ce{Cu2OSeO3} nanoparticles can be produced that have a very narrow spread of sizes, or nanoparticles can be easily separated by size, there is potential for further experiments to investigate the magnetic structure - nanoparticle size relationship.
Investigation should also be conducted into nanoparticles of other skyrmion hosting materials, such as FeGe \cite{yu2011near} and Co-Zn-Mn \cite{tokunaga2015new}, which are simpler systems to study the effects of confinement in nanoparticles.
In addition, other materials which host different types of topological magnetism such as N\'eel \cite{kezsmarki2015neel, bordacs2017equilibrium, ruff2015multiferroicity} and anti-skyrmions \cite{nayak2017magnetic} provide a fruitful route for further investigations.

\section*{Acknowledgements}
This work is financially supported by the EPSRC UK Skyrmion Project Grant EP/N032128/1.
We would like to thank Steve York (University of Warwick) for assistance with the SEM measurements and David W. Lester (University of Warwick) for help with the laser diffraction measurements.

\section*{References}

\end{document}